\documentclass[twocolumn,prl,floatfix,superscriptaddress,nobibnotes]{revtex4}

\usepackage{verbatim}
\usepackage{amsmath}
\usepackage{amssymb}
\usepackage{graphicx}
\usepackage{hyperref}
\usepackage{multirow}
\usepackage{array}
\usepackage{float}

\usepackage{enumitem}
\usepackage{dcolumn}
\usepackage{bm}
\usepackage{verbatim}
\usepackage{multirow}

\begin{document}

\title{Parity lifetime of bound states in a proximitized semiconductor nanowire}

\author{A.~P.~Higginbotham}

\thanks{These authors contributed equally to this work.}
\affiliation{Center for Quantum Devices, Niels Bohr Institute, University of Copenhagen, Universitetsparken 5, 2100 Copenhagen \O, Denmark}
\affiliation{Department of Physics, Harvard University, Cambridge, Massachusetts 02138, USA}

\author{S.~M.~Albrecht}

\thanks{These authors contributed equally to this work.}
\affiliation{Center for Quantum Devices, Niels Bohr Institute, University of Copenhagen, Universitetsparken 5, 2100 Copenhagen \O, Denmark}

\author{G.~Kir{\v{s}}anskas}

\affiliation{Center for Quantum Devices, Niels Bohr Institute, University of Copenhagen, Universitetsparken 5, 2100 Copenhagen \O, Denmark}

\author{W.~Chang}
\affiliation{Center for Quantum Devices, Niels Bohr Institute, University of Copenhagen, Universitetsparken 5, 2100 Copenhagen \O, Denmark}
\affiliation{Department of Physics, Harvard University, Cambridge, Massachusetts 02138, USA}

\author{F.~Kuemmeth}

\affiliation{Center for Quantum Devices, Niels Bohr Institute, University of Copenhagen, Universitetsparken 5, 2100 Copenhagen \O, Denmark}

\author{P.~Krogstrup}

\affiliation{Center for Quantum Devices, Niels Bohr Institute, University of Copenhagen, Universitetsparken 5, 2100 Copenhagen \O, Denmark}

\author{T.~S.~Jespersen}

\affiliation{Center for Quantum Devices, Niels Bohr Institute, University of Copenhagen, Universitetsparken 5, 2100 Copenhagen \O, Denmark}

\author{J.~Nyg{\aa}rd}

\affiliation{Center for Quantum Devices, Niels Bohr Institute, University of Copenhagen, Universitetsparken 5, 2100 Copenhagen \O, Denmark}

\author{K.~Flensberg}

\affiliation{Center for Quantum Devices, Niels Bohr Institute, University of Copenhagen, Universitetsparken 5, 2100 Copenhagen \O, Denmark}

\author{C.~M.~Marcus}

\affiliation{Center for Quantum Devices, Niels Bohr Institute, University of Copenhagen, Universitetsparken 5, 2100 Copenhagen \O, Denmark}

\date{\today}

\maketitle

\textbf{
Quasiparticle excitations can compromise the performance of superconducting devices, causing high frequency dissipation,
decoherence in Josephson qubits \cite{Lang:2003fk,Aumentado:2004ij,Martinis:2009bd,DeVisser:2011bm,Saira:2012ca,Pop:2014gg}, and braiding errors in proposed Majorana-based  topological quantum computers \cite{Leijnse:2011dfa,Rainis:2012uw,Cheng:2012bu}. Quasiparticle dynamics have been studied in detail in metallic superconductors \cite{Ferguson:2006hg,Zgirski:2011wp,Sun:2012gl,Riste:2013il,Maisi:2013cr}
but remain relatively unexplored in semiconductor-superconductor structures, which are now being intensely pursued  in the context of topological superconductivity.
To this end, we introduce a new physical system comprised of a gate-confined semiconductor nanowire with an epitaxially grown superconductor layer, yielding an isolated, proximitized nanowire segment. We identify Andreev-like bound states in the semiconductor via bias spectroscopy, determine the characteristic temperatures and magnetic fields for quasiparticle excitations, and extract a parity lifetime (poisoning time) of the bound state in the semiconductor exceeding 10~ms.
}

Semiconductor-superconductor hybrids have been investigated for many years \cite{Doh:2005wu,Hofstetter:2009bm,Pillet:2010ds,DeFranceschi:2010tq,Giazotto:2011jq}, but recently have received  renewed interest in the context of topological superconductivity, motivated by the realization that combining spin-orbit interaction, Zeeman splitting and proximity coupling to a conventional s-wave superconductor provides the necessary ingredients to create Majorana modes at the ends of a one-dimensional (1D) wire.  Such modes are expected to show nonabelian statistics, allowing, in principle, topological encoding of quantum information  \cite{Sau:2010cl,Lutchyn:2010hp,Oreg:2010gk} among other interesting effects \cite{Padurariu:2010fr,Leijnse:2013ba}.

Transport experiments on semiconductor nanowires proximitized by a grounded superconductor have recently revealed characteristic features of  Majorana modes \cite{Mourik:2012je,Das:2012hi,Deng:2012gn,Churchill:2013cq}.
Semiconductor quantum dots with superconducting leads have also been explored experimentally \cite{Deacon:2010jna,Lee:2012ft,Lee:2013gj,Chang:2013fo}, and have been proposed as a basis for Majorana chains \cite{Sau:2012iu,Leijnse:2012jf,Fulga:2013dd}.
Here, we expand the geometries investigated in this context by creating an isolated semiconductor-supercondutor hybrid quantum dot (HQD) connected to normal leads. The device forms the basis of an isolated Majorana system with protected total parity, where both the semiconductor nanowire and the metallic superconductor are mesoscopic \cite{Fu:2010ho,Hutzen:2012gg}.

\begin{figure}[b]
\center
\includegraphics[width = 7.5 cm]{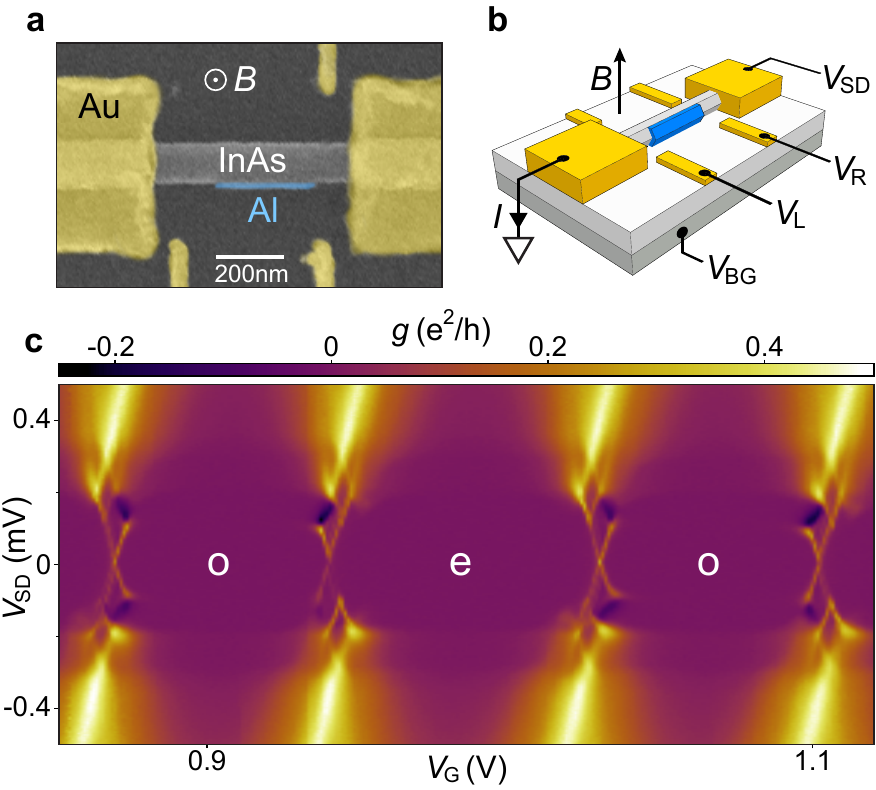}
\caption{\footnotesize{\textbf{Nanowire-based hybrid quantum dot.} \textbf{a}, Scanning electron micrograph of the reported device, consisting of an InAs nanowire (gray) with segment of epitaxial Al on two facets (blue) and Ti/Au contacts and side gates (yellow) on a doped silicon substrate.
\textbf{b}, Device schematic and measurement setup, showing orientation of magnetic field, $B$.
\textbf{c}, Differential conductance, $g$,  as a function of effective gate voltage, $V_\text{G}$, and source-drain voltage, $V_{\text{SD}}$, at $B=0$.  Even (e) and odd (o) occupied Coulomb valleys labeled.
}}
\label{fig1}
\end{figure}

The measured device consists of an InAs nanowire with epitaxial superconducting Al on two facets of the hexagonal wire, with Au ohmic contacts (Figs.~1a,b). Four devices showing similar behavior have been measured. 
The InAs nanowire was grown without stacking faults using molecular beam epitaxy with Al deposited \textit{in situ} to ensure high-quality proximity effect \cite{Krogstrup:2015en,Chang:2015kw}.
Differential conductance, $g$, was measured in a dilution refrigerator with base electron temperature $T\sim50~\mathrm{mK}$ using standard ac lock-in techniques. 
Local side gates, patterned with electron beam lithography, and a global back gate were adjusted to form an Al-InAs HQD in the Coulomb blockade regime, with gate-controlled weak tunneling to the leads.
The lower right gate, $V_\mathrm{R}$, was used to tune the occupation of the dot, with a linear compensation from the lower left gate, $V_\mathrm{L}$, to keep tunneling to the leads symmetric.
We parameterize this with a single effective gate voltage, $V_\mathrm{G}$ (see Supplement).

Differential conductance as a function of $V_\mathrm{G}$ and source-drain bias, $V_\mathrm{SD}$, reveals a series of Coulomb diamonds, corresponding to incremental single-charge states of the HQD (Fig.~1c).
While conductance features at high bias are essentially identical in each diamond, at low bias, $V_\mathrm{SD} <  $ 0.2 mV, a distinctive even-odd pattern of left- and right-facing conductance features is observed. This results in an even-odd alternation of Coulomb blockade peak spacings at zero bias, similar to even-odd spacings seen in metallic superconductors \cite{Tuominen:1992ip,Lafarge:1993vn}. However, the parity-dependent reversing pattern of subgap features at nonzero bias has not been reported before, to our knowledge.
The even-odd pattern indicates that a parity-sensitive bound state is being filled and emptied as electrons are added to the HQD. 

Measured charging energy, $E_C = 1.1~\mathrm{meV}$, and superconducting gap, $\Delta=180~\mathrm{ \mu eV}$, satisfy the condition ($\Delta\ <\ E_C$) for single electron charging \cite{Averin:1992hy,Matveev:1994fq}.
Differential conductance at low bias occurs in a series of narrow features symmetric about zero bias, suggesting transport through an Andreev-like bound state, with negative differential conductance (NDC) observed at the border of odd diamonds.
NDC arises from slow quasiparticle escape, as discussed below, similar to current-blocking seen in metallic superconducting islands in the opposite regime, $\Delta > E_C$ \cite{Hekking:1993gc,Hergenrother:1994ei}.

\begin{figure}[t!]
\center
\includegraphics[]{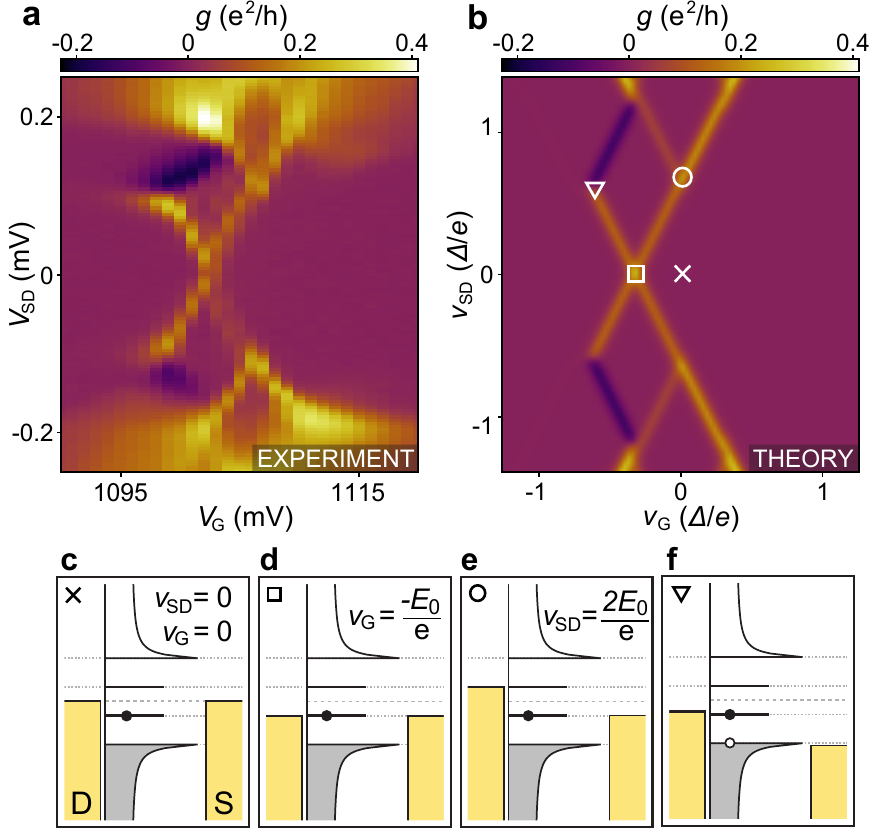}
\caption{\footnotesize{\textbf{Subgap bias spectroscopy, experiment and model.}
\textbf{a}, Experimental differential conductance, $g$, as a function of gate voltage $V_{\mathrm{G}}$ and source-drain $V_\mathrm{SD}$, shows characteristic pattern including negative differential conductivity (NDC).
\textbf{b}, Transport model of \textbf{a}.
$v_\text{G} = \alpha V_\mathrm{G}$ up to an offset, where $\alpha$ is the gate lever arm.
Axis units are $\Delta/e=180~\mathrm{\mu V}$, where $\Delta$ is the superconducting gap.
 See text for model parameters.
\textbf{c}, Source and drain (gold) chemical potentials align with the middle of the gap in the HQD density of states.
No transport occurs due to the presence of superconductivity.
\textbf{d}, Discrete state in resonance with the leads at zero bias.
Transport occurs through single quasiparticle states.
\textbf{e}, Discrete state in resonance with the leads at high bias.
Transport occurs through single and double (particle-hole) quasiparticle states.
\textbf{f}, Discrete state and BCS continuum in the bias window.
Transport is blocked when a quasiparticle is in the continuum, resulting in NDC.
}}
\label{fig2}
\end{figure}

To gain quantitative understanding of these features, we model transport through a single Andreev bound state in the InAs plus a Bardeen-Cooper-Schriffer (BCS) continuum in the Al.
The model assumes symmetric coupling of both the bound state and continuum to the leads, motivated by the  observed symmetry in $V_{SD}$ of the Coulomb diamonds.
Transition rates were calculated from Fermi's golden rule and a steady-state Pauli master equation was solved for state occupancies.
Conductance was then calculated from occupancies and transition rates (see Supplement). 

\begin{figure}
\center \label{figure3}
\includegraphics[scale=1]{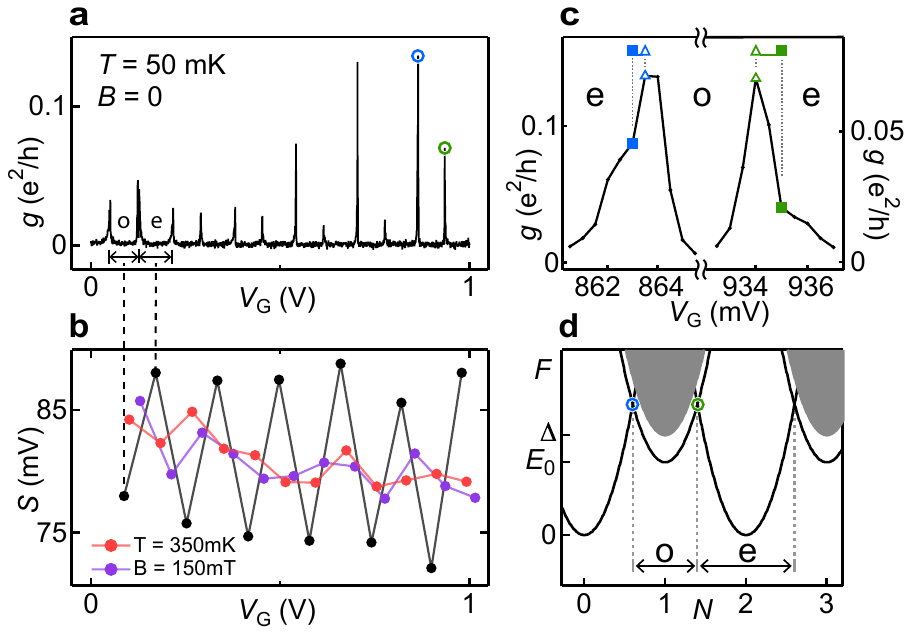}
\caption{\footnotesize{\textbf{Even-odd Coulomb peak spacings}
\textbf{a}, Measured zero-bias conductance, $g$, versus gate voltage, $V_\mathrm{G}$, at temperature $T\sim50~\mathrm{mK}$, and magnetic field $B=0$.
\textbf{b}, Peak spacing, $S$, versus gate voltage. Black points show spacings from \textbf{a} calculated using the peak centroid (first moment), red points $T=350~\mathrm{mK}$ and $B=0$, purple points $B=150~\mathrm{mT}$ and $T\sim50~\mathrm{mK}$.
\textbf{c}, Right-most peaks in \textbf{a}.
Peak maxima ($\triangle$) and centroids ($\blacksquare$) are marked.
\textbf{d}, Free energy, $F$, at $T=0$ versus gate-induced charge, $N$, for different HQD occupations, where $N = C V_\mathrm{G}/e$ up to an offset and $C$ is the gate capacitance.
Parabola intersection points are indicated by circles, corresponding to Coulomb peaks.
BCS continuum (shaded), shown for odd occupancy.
Odd Coulomb diamonds carry an energy offset $E_0$ for quasiparticle occupation of the sub gap state, resulting in a difference in spacing for even and odd diamonds. 
}}
\end{figure}

Measured and model conductances are compared in Figs.~2a,b.
The coupling of the bound state to each lead, noting the near-symmetry of the diamonds, was estimated to be $\Gamma_0 = 0.5~\mathrm{GHz}$, based on zero-bias conductance (Fig.~2d).
The energy of the discrete state, $E_0=58~\mathrm{\mu e V}$ at zero magnetic field, was measured using finite bias spectroscopy (Fig.~2e).
The normal-state conductance from each lead to the continuum, $g_\mathrm{Al} = 0.15~e^2/h$, was estimated by comparing Coulomb blockaded transport features in the high bias regime ($V_\mathrm{SD}=0.4~\mathrm{mV}$). 
The superconducting gap, $\Delta=180~\mathrm{\mu eV}$, was found from the onset of NDC, which is expected to occur at $e V_\mathrm{SD}=\Delta-E_0$  (Fig.~2f).
While the rate model shows good agreement with experimental data, some features are not captured, including broadening at high bias, with greater broadening correlated with weaker NDC, and peak-to-peak fluctuations in the slope of the NDC feature.
These features may be related to heating or cotunneling, not accounted for by the model.

\begin{figure*}[t]
\center \label{figure4}
\includegraphics[]{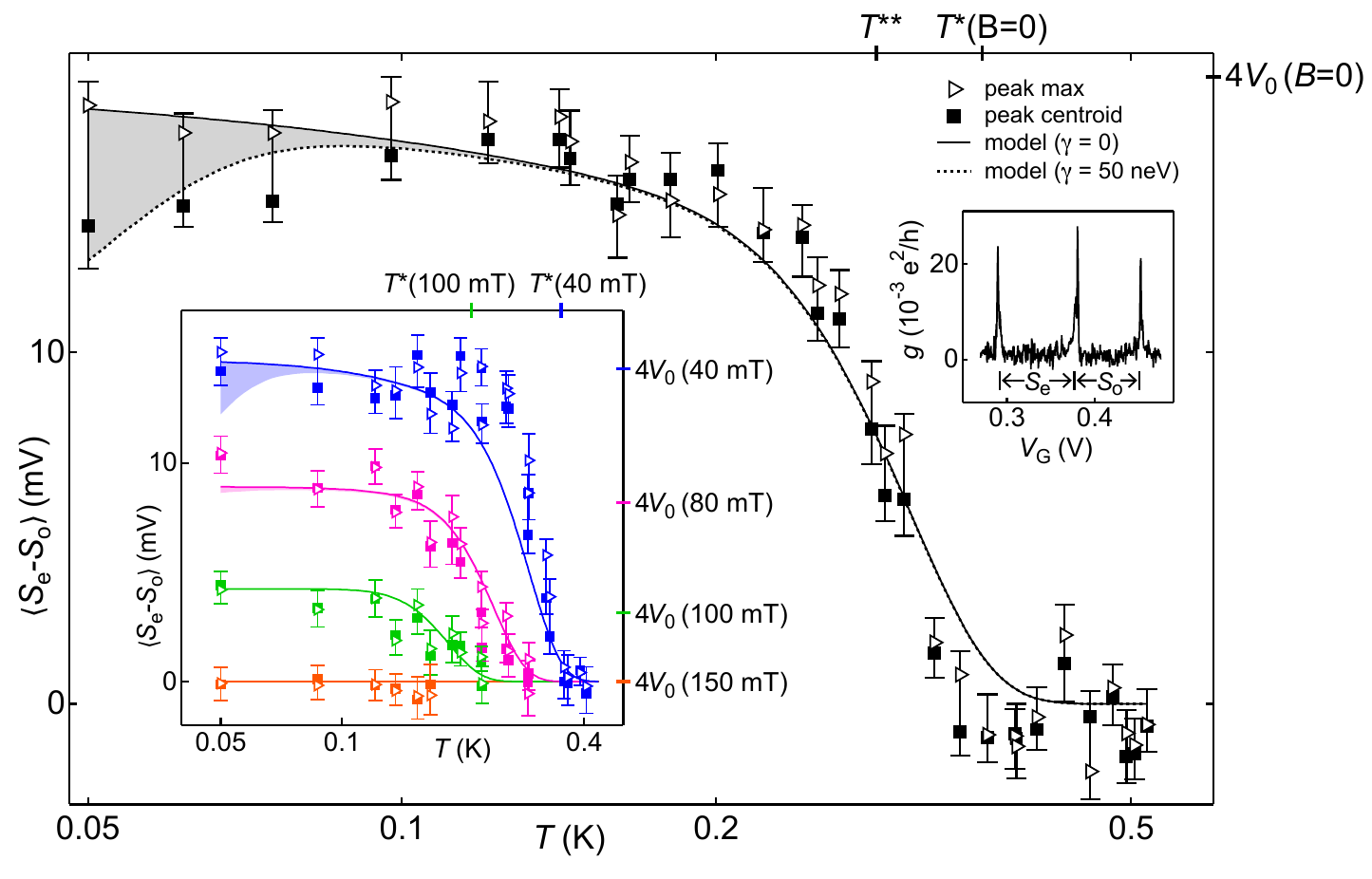}
\caption{\footnotesize{\textbf{Temperature and magnetic field dependence of the even-odd peak spacings.}
Average even-odd spacing difference, $\left< S_\mathrm{e} - S_\mathrm{o} \right>$, versus temperature, $T$.
Spacing between peak maxima (triangle) and centroids (square) are shown.
Spacing expected from lower Zeeman-split bound state, $4 V_0( B ) = 4 E_0( B ) / (\alpha e)$, indicated on right axis.
Quasiparticle activation temperature, $T^*$, and crossover temperature, $T^{**}$, indicated on top axis.
Solid curve is Eq.~(\ref{eq:seo}) with a HQD density of states measured from Fig.~2 ($\Delta=180~\mathrm{\mu e V}$, $E_0=58~\mathrm{\mu e V}$, $\alpha=0.013$), and the fitted aluminum volume, $V=7.4 \times 10^4~\mathrm{nm^3}$.
Dotted curve includes a discrete state broadening, $\gamma=50~\mathrm{neV}$, fit to the centroid data.
\textit{Left inset:} Same as main, but at $B = 40, 80, 100, 150~\mathrm{mT}$, from top to bottom.
Curves are fit to two shared parameters: g-factor, $g=6$, and superconducting critical field, $B_\mathrm{c} = 120~\mathrm{mT}$, with other parameters fixed from main figure.
\textit{Right inset:} Representative Coulomb peaks showing even ($S_\mathrm{e}$) and odd ($S_\mathrm{o}$) spacings.
}}
\end{figure*}

The observation of negative differential conductance  places a bound on the relaxation rate of a single quasiparticle in the HQD from the continuum (in the Al) to the bound state (in the InAs nanowire).
Negative differential conductance arises when an electron tunnels into the weakly coupled BCS continuum, blockading transport until it exits via the lead. The blocking condition is shown for a hole-like excitation in Fig.~2f.
Unblocking occurs when the quasiparticle relaxes into the bound state, followed by a fast escape to the leads.
NDC thus indicates a long quasiparticle relaxation time, $\tau_\mathrm{qp}$, from the continuum to the bound state. Using independently determined parameters, the observed NDC is only compatible with the model when $\tau_{\text{qp}} > 0.1~\mathrm{\mu  s}$ (see Supplement). This bound on $\tau_{\text{qp}}$ is used below to similarly constrain the characteristic poisoning time for the bound state.

Turning our attention to the even-odd structure of zero-bias Coulomb peaks (Figs.~3a,b), we observed consistent large-small peak spacings (Fig.~3), associating the larger spacings with even occupation, as expected theoretically \cite{Averin:1992hy,Matveev:1994fq} and already evident in Fig.~1. Occasional even-odd parity reversals on the timescale of hours were observed in some devices, similar to what is seen in metallic devices \cite{Maisi:2013cr}. Peak spacing alternation disappears at higher magnetic fields, $B$, consistent with the superconducting-to-normal transition, and also disappears at elevated temperature, $T>0.4~\mathrm{K}$, significantly below the superconducting critical temperature, $T_c \sim 1~\mathrm{K}$. The temperature dependence is consistent with similar behavior seen in metallic structures \cite{Tuominen:1992ip,Lafarge:1993vn}, and can be understood as the result of thermal activation of quasiparticles within the HQD with fixed total charge.

As seen in Fig.~3c, individual Coulomb peaks are asymmetric in shape, with their centroids (first moments) on the even sides of the peak maxima. Note that the asymmetry leads to higher near-peak conductance in {\it even} valleys, the opposite of the Kondo effect.
The asymmetric shape is most pronounced at low temperature, $T<0.15~\mathrm{K}$, and decreases with increasing magnetic field. 
The degree of asymmetry is not predicted by the rate model, even taking into account the known small asymmetry due to spin degeneracy \cite{Aleiner:2002in}.
In the analysis below, we consider peak positions defined both by peak maxima and centroids.

A model of even-odd Coulomb peak spacing that includes thermal quasiparticle excitations follows earlier treatments \cite{Tuominen:1992ip,Lafarge:1993vn,Matveev:1994fq}, including a discrete subgap state as well as the BCS continuum \cite{Lafarge:1993vn} (Fig.~3d).
Even-odd peak spacing difference, $S_\mathrm{e} - S_\mathrm{o}$, depends on the difference of free energies,\begin{equation}
\label{eq:seo}
S_\mathrm{e} - S_\mathrm{o}  = \frac{4}{\alpha e} \left( F_\mathrm{o} - F_\mathrm{e} \right),
\end{equation}
where $\alpha$ is the (dimensionless) gate lever arm.
The free energy difference, written in terms of the ratio of partition functions,
\begin{equation}
\label{eq:feo}
F_\mathrm{o}-F_\mathrm{e} = -k_\mathrm{B} T \ln \left( \frac{ Z_\mathrm{o} }{ Z_\mathrm{e} } \right),
\end{equation} 
depends on $D(E)$, the density of states of the HQD, 
\begin{equation}
\frac{Z_\mathrm{o}}{Z_\mathrm{e}} = \int_0^\infty \! \mathrm{d} E \ D( E ) \ln \coth[E / (2 k_{\mathrm{B}}T )],
\end{equation}
where $D(E)$ consists of one subgap state and the continuum. For $\Delta \gg k_{\mathrm{B}}T$, this can be written
\begin{equation}
F_\mathrm{o}-F_\mathrm{e} \approx -k_\mathrm{B} T \ln( N_\mathrm{eff} e^{-\Delta / k_{\mathrm{B}} T} + 2 e^{ - E_0 / k_{\mathrm{B}} T }),
\end{equation}
where $N_\mathrm{eff} = \rho_\mathrm{Al} V \sqrt{ 2 \pi k_{\mathrm{B}} T \Delta }$ is the effective number of continuum states for Al volume, $V$, and normal density of states $\rho_\mathrm{Al}$ \cite{Tuominen:1992ip,Lafarge:1993vn} (see Supplement).

Within this model, one can identify a characteristic temperature, $T^* \sim \Delta / [ k_\mathrm{B} \ln( N_\mathrm{eff} )]$, less than the gap, above which even-odd peak spacing alternation is expected to disappear.
Note in this expression $N_\mathrm{eff}$ itself depends on $T$, and also that $T^{*}$ does not depend on the bound state energy, $E_{0}$.
A second (lower) characteristic temperature, $T^{**} \sim ( \Delta - E_0 ) /[ k_\mathrm{B} \ln ( N_\mathrm{eff}/2 ) ]$, which does depend on $E_{0}$, is where the even-odd alternation is affected by the bound state, leading to saturation at low temperature \cite{Tuominen:1992ip,Lafarge:1993vn}.
For a spin-resolved zero-energy ($E_{0} =0$) bound state---the case for unsplit Majorana zero modes---these characteristic temperatures coincide and even-odd structure vanishes, as pointed out in Ref.~\cite{Fu:2010ho}. In the opposite case, where the bound state reaches the continuum ($E_{0} =\Delta$), the saturation temperature vanishes, $T^{**} =0$, and the metallic result with no bound state is recovered \cite{Tuominen:1992ip,Lafarge:1993vn}.

Experimentally, the average even-odd peak spacing difference, $\langle S_\mathrm{e}-S_\mathrm{o} \rangle$, was determined by averaging over a set of 24 consecutive Coulomb peak spacings, including those shown in Fig.~3, at each temperature.
Figure 4 shows even-odd peak spacing difference appearing abruptly at $T_{\mathrm{onset}}  \sim 0.4$~K, and saturating at $T_{\mathrm{sat}} \sim 0.2$~K, with a saturation amplitude near the value expected from the measured bound state energy, $4 V_0 = 4 E_{0}/(\alpha e) $.
Figure 4 shows good agreement between experiment and the model, Eq.~(\ref{eq:seo}), using a density of states determined independently from data in Fig.~2,  with $V=7.4 \times 10^4~\mathrm{nm^3}$ as a fit parameter, consistent with the micrograph (Fig.~1a), and $\rho_{\mathrm{Al}} = 23~\mathrm{eV^{-1} nm^{-3}}$ \cite{Maisi:2013cr}.

The asymmetric peak shape complicates measurement of even-odd spacings, as one can either use the centroids or maxima to measure spacings, the two methods giving different results. 
Larger peak tails on the even valley side cause the centroids to be more regularly spaced than the maxima. 
This is evident in Fig.~4, where the centroid method shows a decreasing peak spacing difference at low temperature, while with the maximum method the spacing remains flat.
The thermal model of $S_\mathrm{e}-S_\mathrm{o}$ can also show a decrease at low temperature if broadening of the bound state is included  (See Methods).
We do not understand at present if the low temperature decrease in the centroid data is related to the decrease seen in the model when broadening is included.
It is worth noting, however, that the fit to the centroid data gives a broadening $\gamma=50~\mathrm{neV}$, reasonably close to the value estimated from the lead couplings, $(h \Gamma_0)^2/\Delta = 20~\mathrm{neV}$.

Applied magnetic field (direction shown in Fig.~1b) reduces the characteristic temperatures $T_{\mathrm{onset}}$, $T_{\mathrm{sat}}$, and  saturation amplitudes.
Field dependence is modeled by including Zeeman splitting of the bound state and orbital reduction of the gap and bound state energy, taking the $g$-factor and critical magnetic field as two fit parameters applied to all data sets.
The fit value $g=6$ lies within the typical range for InAs nanowires \cite{Bjork:2005hx,Csonka:2008gz}, supporting our interpretation of the bound state residing in the InAs. The fit value of critical field, $B_\mathrm{c}=120 ~\mathrm{mT}$, is typical for this geometry.

Good agreement between the peak spacing data and the thermodynamic model (Fig.~4) suggests that the number of thermally activated quasiparticles obeys equilibrium statistics, $N_\mathrm{eq}(T) = N_\mathrm{eff}^2 e^{- 2 \Delta / k_\mathrm{B} T}$ (see Supplement for derivation). Saturation caused by the bound state means that even-odd amplitude loses sensitivity as a quasiparticle detector below $T_{\mathrm{sat}}$. 
We therefore take $N_\mathrm{eq}( T_\mathrm{sat}) \sim 10^{-5}$ (for $T_\mathrm{sat} \sim 0.2$ K) as an upper bound for the number of quasiparticles at temperatures below $T_\mathrm{sat}$.
The corresponding upper bound of the quasiparticle fraction, $x_\mathrm{qp} = N_\mathrm{eq}( T_\mathrm{sat} ) / ( \rho_\mathrm{Al} V \Delta )\sim 10^{-8}$, is comparable to values in the recent literature, $10^{-5}-10^{-8}$, for metallic superconducting junctions and qubits~\cite{Martinis:2009bd,DeVisser:2011bm,Saira:2012ca,Riste:2013il,Pop:2014gg}.

We now discuss the implications of our measurements for determining the poisoning time, $\tau_\mathrm{p}$, of the bound state. For the present geometry, the dominant source of poisoning of the bound state is not tunneling of electrons from the leads, which is negligible in the strongly blockaded regime, but is rather the continuum in the strongly-coupled Al, within the isolated structure itself.
Theoretical estimates \cite{Rainis:2012uw} suggest an inverse relationship between $\tau_{p}$  and the number of available quasiparticles, with a proportionality that depends on system details.
Taking the bound on single quasiparticle relaxation time from the continuum into the bound state, $\tau_{\text{qp}} > 0.1~\mathrm{\mu  s}$, from above, as the poisoning time when a single quasiparticle is present, we estimate $\tau_p$ by scaling for the actual number of quasiparticles in equilibrium, $N_\mathrm{eq}$, giving a poisoning time $\tau_\mathrm{p} = \tau_\mathrm{qp} / N_\mathrm{eq} \gtrsim 10~\mathrm{ms}$.

We expect $\tau_{p}$ to depend weakly on the bound state energy for low-energy bound states \cite{Zgirski:2011wp,Olivares:2014ig,Zazunov:2014kn}, including for Majorana zero modes at $E_0 = 0$.
Device geometry may somewhat alter the number of quasiparticles available to relax into the bound state, \textit{i.e.} by changing $N_\mathrm{eff}$, but any increase can be compensated by exponentially small decreases in the quasiparticle temperature.
The long poisoning time obtained here suggests that a large number of braiding operations in Majorana systems should be readily achievable within the relevant time scale.

\textbf{Methods}

\textit{ Sample preparation: } InAs nanowires were grown in the [001] direction with wurzite crystal structure with Al epitaxially matched to [111] on two of the six  $\{1\bar{1}00\}$ sidefacets.
They were then deposited randomly onto a doped silicon substrate with 100 nm of thermal oxide.
 Electron-beam lithographically patterned wet etch of the epitaxial Al shell (Transene Al Etchant D, 55 C, 10 s) resulted in a submicron Al segment (310 nm, Fig.~\ref{fig1}a). Ti/Au (5/100 nm) ohmic contacts were deposited on the ends following {\em in situ} Ar milling (1  mTorr, 300 V, $75~\mathrm{s}$), with side gates deposited in the same step. For the present device, the end of the upper left gate broke off during processing. However, the device could be tuned well without it.

\textit{ Master equations: }
The master equations (used for Fig.~1b) consider states with fixed total parity, composed of the combined parity of quasiparticles in the thermalized continuum and the 0, 1, or 2 quasiparticles in the bound state (see Supplement).

\textit{ Free energy model: } 
Even and odd partition functions in Eq.~2, $F_\mathrm{o}-F_\mathrm{e} = -k_\mathrm{B} T \ln( Z_\mathrm{o} / Z_\mathrm{e} )$, can be written as sums of Boltzmann factors over respectively odd and even occupancies of the isolated island.
For even-occupancy,
\begin{eqnarray}
Z_\mathrm{e} = 1 + \sum_{i \neq j}  e^{-E_i/ k_\mathrm{B} T} e^{-E_j/ k_\mathrm{B} T} + ...,
\end{eqnarray}
where the first term stands for zero quasiparticles, the second for two (at energies $E_i$ and $E_j$), and additional terms for four, six, etc.
$Z_\mathrm{o}$ similarly runs over odd occupied states.
Rewriting these sums as integrals over positive energies yields
\begin{equation}
\label{eq:eint}
F_\mathrm{o} - F_\mathrm{e} = -k_\mathrm{B} T \ln \tanh  \int_0^\infty \! \mathrm{d} E \ D( E ) \ln \coth( E / 2k_\mathrm{B}T ),
\end{equation}
where $D( E )$ is the density of states of the HQD,
\begin{equation}
D(E) = \rho_\mathrm{BCS}( E ) + \frac{1}{2} \rho_\mathrm{0}^{+}( E ) + \frac{1}{2} \rho_\mathrm{0}^{-}( E ).
\end{equation}
We take $\rho_\mathrm{BCS}( E )$ to be a standard BCS density of states,
\begin{equation}
\rho_\mathrm{BCS}( E ) = \frac{ \rho_\mathrm{Al} V  E }{\sqrt{ E^2 - \Delta(B)^2 } } \theta( E - \Delta )
\end{equation}
($\theta$ is the step function), and $\rho_0$ to be a pair of Lorentzian-broadened spinful levels symmetric about zero,
\begin{equation}
\rho_0^{\pm}( E ) = \frac{ \gamma/ 2 \pi}{ (E-E^{\pm}_0 )^2 + (\gamma/2)^2} + \frac{ \gamma / 2 \pi }{ (E+E^{\pm}_0 )^2 + (\gamma/2)^2}.
\end{equation}
Zeeman splitting of the bound state and pair-breaking by the external magnetic field are modeled with the equations
\begin{eqnarray}
E_0^{\pm}(B) &=& \frac{\Delta(B)}{\Delta} E_0 \pm \frac{1}{2} g \mu_B B, \label{eq:epm} \\
\Delta( B ) &=& \Delta \sqrt{ 1 - \left( \frac{B}{B_c} \right)^2 } \label{eq:deltab},
\end{eqnarray}
where $E_\mathrm{0}$ is the zero-field state energy and $\Delta$ is the zero field superconducting gap.
In the event that a bound state goes above the continuum, $E_s^{+} > \Delta( B )$, we no longer include the state in the free energy.
Equation~(\ref{eq:eint}) was integrated numerically to obtain theory curves in Fig.~(4).

Equations (\ref{eq:epm}) and (\ref{eq:deltab}) are reasonable provided the lower spin-split state remains at positive energy, $E_0^- > 0$. For sufficiently large $B_c$, the bound state will reach zero energy, resulting in topological superconductivity and Majorana modes, the subject of future work.

\vspace{0.1in}

We thank Leonid Glazman, Bert Halperin, Roman Lutchyn and Jukka Pekola for valuable discussions, and Giulio Ungaretti, Shivendra Upadhyay and Claus S{\o}rensen for contributions to growth and fabrication. Research support by Microsoft Project Q, the Danish National Research Foundation, the Lundbeck Foundation, the Carlsberg Foundation, and the European Commission. APH acknowledges support from the US Department of Energy, CMM acknowledges support from the Villum Foundation.

\bibliographystyle{naturemag}

\clearpage
\onecolumngrid
\setcounter{figure}{0}
\setcounter{equation}{0}
\setcounter{page}{1}

\renewcommand{\thefigure}{S\arabic{figure}}
\renewcommand{\theequation}{S\arabic{equation}}
\renewcommand{\bibnumfmt}[1]{[S#1]}
\renewcommand{\citenumfont}[1]{S#1}

\renewcommand{\figurename}{Fig.}
\renewcommand{\tablename}{Table}
\newcommand{\sectionname}{Section}

\newcommand{\e}{e}
\renewcommand{\i}{i}
\newcommand{\m}{\mathcal}
\newcommand{\ve}{\varepsilon}
\newcommand{\vphi}{\varphi}
\newcommand{\pd}{\partial}
\newcommand{\dif}[1]{\mathrm{d}#1\,}

\newcommand\ph[1]{\phantom{#1}}
\newcommand\abs[1]{\lvert#1\rvert}
\newcommand\absB[1]{\left\lvert#1\right\rvert}
\newcommand\avg[1]{\left\langle#1\right\rangle}
\newcommand\avgs[1]{\langle#1\rangle}
\newcommand\bras[1]{\left\langle#1\right\rvert}
\newcommand\kets[1]{\left\lvert#1\right\rangle}
\newcommand\bra[1]{\langle#1\rvert}
\newcommand\ket[1]{\lvert#1\rangle}
\newcommand\braket[2]{\langle#1\rvert#2\rangle}
\newcommand{\tg}{\operatorname{tg}}
\newcommand{\ctg}{\operatorname{ctg}}
\newcommand{\arctg}{\operatorname{arctg}}
\newcommand{\Real}{\operatorname{Re}}
\newcommand{\Imag}{\operatorname{Im}}
\newcommand{\Realp}{\operatorname{Re}\hspace{0.025cm}}
\newcommand{\Imagp}{\operatorname{Im}\hspace{0.025cm}}
\newcommand{\Arg}{\operatorname{Arg}}
\newcommand{\Tr}{\operatorname{Tr}}
\newcommand{\Res}{\operatorname{Res}}
\newcommand{\rot}{\operatorname{rot}}
\newcommand{\dive}{\operatorname{div}}
\newcommand\To{T_\tau}
\newcommand\Li{\operatorname{Li}}
\newcommand\sgn{\operatorname{sgn}}
\newcommand\dilog{\operatorname{dilog}}
\newcommand\adilog{\operatorname{adilog}}

\newcommand{\up}{\uparrow}
\newcommand{\down}{\downarrow}
\newcommand{\uptld}{\tilde{\uparrow}}
\newcommand{\downtld}{\tilde{\downarrow}}
\newcommand{\upbl}{{\color{blue}\uparrow}}
\newcommand{\downbl}{{\color{blue}\downarrow}}
\newcommand{\etabl}{{\color{blue}\eta}}
\newcommand{\upgr}{{\color{green}\uparrow}}
\newcommand{\downgr}{{\color{green}\downarrow}}
\newcommand{\sigmagr}{{\color{green}\sigma}}

\newcommand{\bkap}{\mathbf{\kappa}}
\newcommand{\ba}{\mathbf{a}}
\newcommand{\bff}{\mathbf{f}}
\newcommand{\bj}{\mathbf{j}}
\newcommand{\bk}{\mathbf{k}}
\newcommand{\bn}{\mathbf{n}}
\newcommand{\br}{\mathbf{r}}
\newcommand{\bp}{\mathbf{p}}
\newcommand{\bq}{\mathbf{q}}
\newcommand{\bss}{\mathbf{s}}
\newcommand{\bu}{\mathbf{u}}
\newcommand{\bv}{\mathbf{v}}
\newcommand{\bA}{\mathbf{A}}
\newcommand{\bB}{\mathbf{B}}
\newcommand{\bE}{\mathbf{E}}
\newcommand{\bF}{\mathbf{F}}
\newcommand{\bL}{\mathbf{L}}
\newcommand{\bS}{\mathbf{S}}
\newcommand{\bQ}{\mathbf{Q}}

\newcommand{\mrC}{\mathrm{C}}
\newcommand{\mrD}{\mathrm{D}}
\newcommand{\mrK}{\mathrm{K}}
\newcommand{\mrL}{\mathrm{L}}
\newcommand{\mrM}{\mathrm{M}}
\newcommand{\mrN}{\mathrm{N}}
\newcommand{\mrR}{\mathrm{R}}
\newcommand{\mrS}{\mathrm{S}}
\newcommand{\mrT}{\mathrm{T}}
\newcommand{\mri}{\mathrm{i}}
\newcommand{\mre}{\mathrm{e}}
\newcommand{\mrp}{\mathrm{p}}

\newcommand{\cd}{c^{\dag}}
\newcommand{\can}{c^{\phantom{\dag}}}
\newcommand{\icd}{\hat{c}^{\dag}}
\newcommand{\ian}{\hat{c}^{\phantom{\dag}}}
\newcommand{\tcd}{\tilde{c}^{\dag}}
\newcommand{\tcan}{\tilde{c}^{\phantom{\dag}}}
\newcommand{\ad}{a^{\dag}}
\newcommand{\aan}{a^{\phantom{\dag}}}
\newcommand{\dd}{d^{\dag}}
\newcommand{\dan}{d^{\phantom{\dag}}}
\newcommand{\gamd}{\gamma^{\dag}}
\newcommand{\gaman}{\gamma^{\phantom{\dag}}}

\newcommand{\T}{\mathrm{T}}
\newcommand{\D}{\mathrm{D}}
\newcommand{\LR}{\mathrm{LR}}
\newcommand{\mrEven}{\mathrm{e}}
\newcommand{\mrOdd}{\mathrm{o}}
\newcommand{\s}{s}
\newcommand{\absDelta}{\Delta}
\newcommand{\mcZ}{Z}
\newcommand\fixme[1]{\textit{#1}}

\section{Supplementary Information:}
\section{Quasiparticle parity dynamics in a superconductor-semiconductor hybrid quantum dot}

\vspace{0.1in}

\begin{enumerate}[nolistsep]
\item \label{sec:gates} Effective gate voltage definition 
\item \label{sec:tqp} Bound on the single quasiparticle relaxation time
\item \label{sec:diamonds} Detailed interpretation of Coulomb diamonds
\item \label{sec:model} Derivation of transport and thermal model
\item \label{sec:dFapprox} Comparison of free energy approximations
\item \label{sec:dFBS} Effect of the bound state on the free energy
\end{enumerate}

\vspace{0.2in}

\subsection{ \ref{sec:gates}. Effective gate voltage definition }
We define an effective gate voltage in software to tune the HQD. The physical gate voltages, $V_\mathrm{R}$ and $V_\mathrm{L}$, are related to the effective gate voltage, $V_\mathrm{G}$, by
\begin{align*}
V_\mathrm{R} &= V_{R,0} + \kappa~V_\mathrm{G}, \\
V_\mathrm{L} &= V_{L,0} + \sqrt{1-\kappa^2}~V_\mathrm{G},
\end{align*}
with $\kappa = 0.9997$ and offset voltages $V_{R,0}=-2.41~\mathrm{V}$, $V_{L,0}=-3.96~\mathrm{V}$.

These transformation rules ensure that $V_G^2 = (V_R - V_{R,0})^2 + (V_L - V_{L,0})^2$, so that $V_G$ can be interpreted as the distance from $(V_{R,0},V_{L,0})$ in the $V_R-V_L$ plane.

All measurements are performed at backgate voltage $V_\mathrm{BG} = 2.39~\mathrm{V}$.

\subsection{ \ref{sec:tqp}. Bound on the single quasiparticle relaxation time}
The effect of quasiparticle relaxation is shown in Figs.~S1a-e.
Quasiparticle relaxation results in a disappearance of the negative differential conductance, in combination with the appearance of an extra conductance threshold.
We quantify this observation by introducing the relative conductance ratio
\begin{equation}
R = \frac{ g' + g_\mathrm{NDC} }{ g' - g_\mathrm{NDC} }
\end{equation}
where $g_\mathrm{NDC}$ is the minimum of the negative differential conductance, and $g'$ is the maximum of the extra conductance threshold that appears when $\tau_\mathrm{qp} \rightarrow 0$ (see Fig.~S1i).
The $R$-value is a metric for the relative strength of the negative differential conductance.

Figs.~S1f-j show example conductance traces at constant bias and their associated $R$-values.
The traces show that $R \approx -1$ corresponds to slow quasiparticle relaxation, and $R \approx 1$ corresponds to fast quasiparticle relaxation.

Fig.~S2 shows the $R$-value calculated as a function of single quasiparticle relaxation time, $\tau_\mathrm{qp}$.
Also shown is a measured $R$-value averaged over all negative differential conductance features in Fig.~1 of the main text.
The measured $R$-value is consistent with $\tau_\mathrm{qp} > 0.1~\mathrm{\mu s}$, giving the experimental bound on the single quasiparticle relaxation time.

\begin{figure}[h]
	\includegraphics[width=7in]{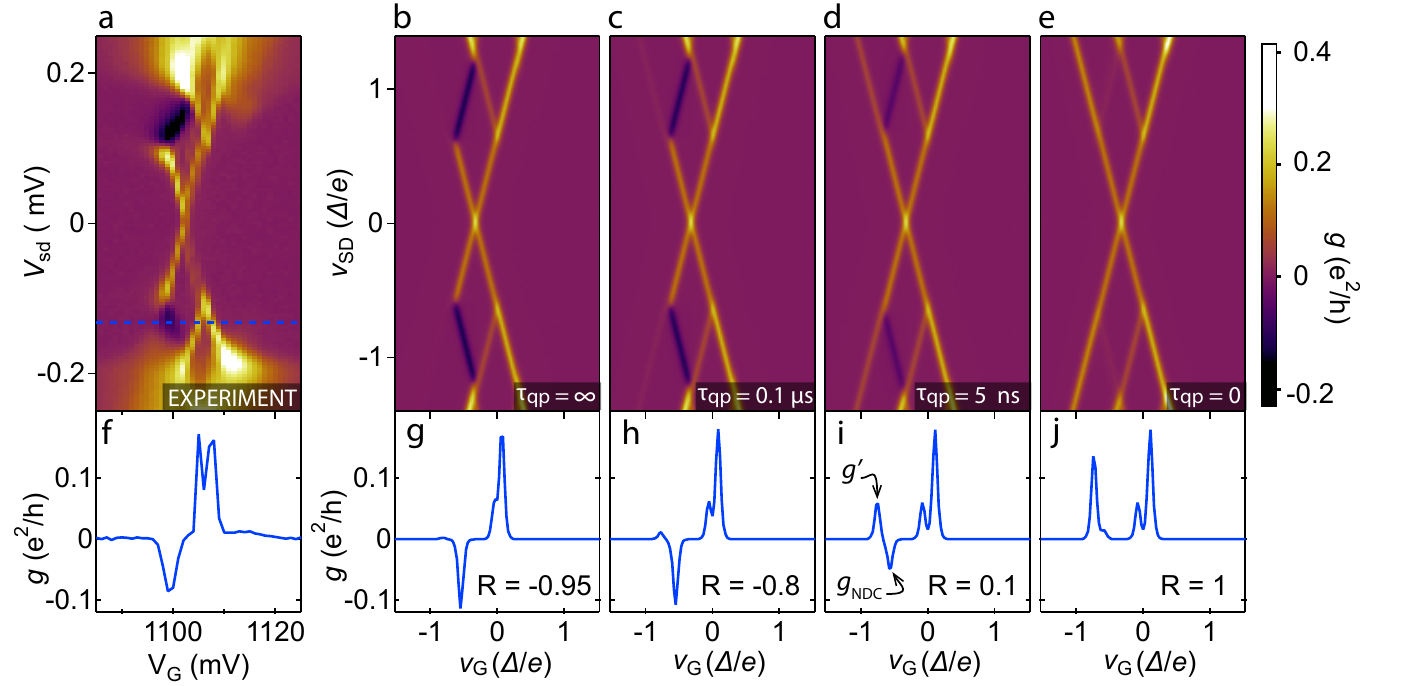}
	\caption{\textbf{Effect of quasiparticle relaxation.} \textbf{a}  Measured conductance $g$ versus source-drain bias $V_\mathrm{SD}$ and gate $V_\mathrm{G}$.
	\textbf{b}, Transport model of \textbf{a}, with $\tau_\mathrm{qp}=\infty$.
$v_\text{G} \equiv \alpha V_\mathrm{G}$ up to an offset, where $\alpha$ is the gate lever arm. Axis units are $\Delta/e=180~\mathrm{\mu V}$.
	\textbf{c-e}, Model with $\tau_\mathrm{qp}=0.1~\mathrm{\mu s}$, $\tau_\mathrm{qp}=5~\mathrm{n s}$, and $\tau_\mathrm{qp}=0$ respectively.
	\textbf{f-j}, Conductance versus gate at constant bias indicated in \textbf{a}. Relative conductance ratio, $R = ( g' + g_\mathrm{NDC} )/(g' - g_\mathrm{NDC} )$, for theory curves is labeled (see text).
	}
\end{figure}

\begin{figure}[h]
	\includegraphics{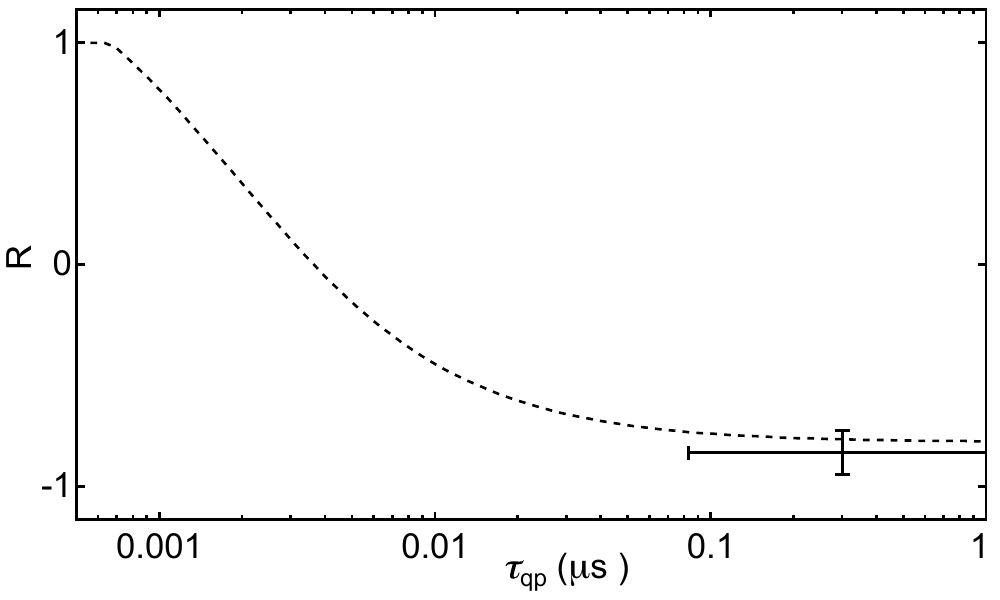}
	\caption{\textbf{Single quasiparticle relaxation bound.} Relative conductance ratio, $R = ( g' + g_\mathrm{NDC} )/(g' - g_\mathrm{NDC} )$, versus single quasiparticle relaxation time $\tau_\mathrm{qp}$.
	Dashed curve is theory derived as shown in Fig.~S1.
	Data is the average over all charge transitions in Fig.~1, with vertical error the standard deviation of the mean, and horizontal error propagated from vertical.
	}
\end{figure}

\subsection{ \ref{sec:diamonds}. Detailed interpretation of Coulomb diamonds}
Each conductance threshold in the Coulomb diamond plots can be interpreted with the help of the transport model, as shown in Fig.~S3.
For example, the highest bias at which NDC is observed occurs at the intersection of black and green lines, when $v_\mathrm{SD}= ( \Delta + E_0 )/e$.

\begin{figure}[h]
\begin{center}
\includegraphics{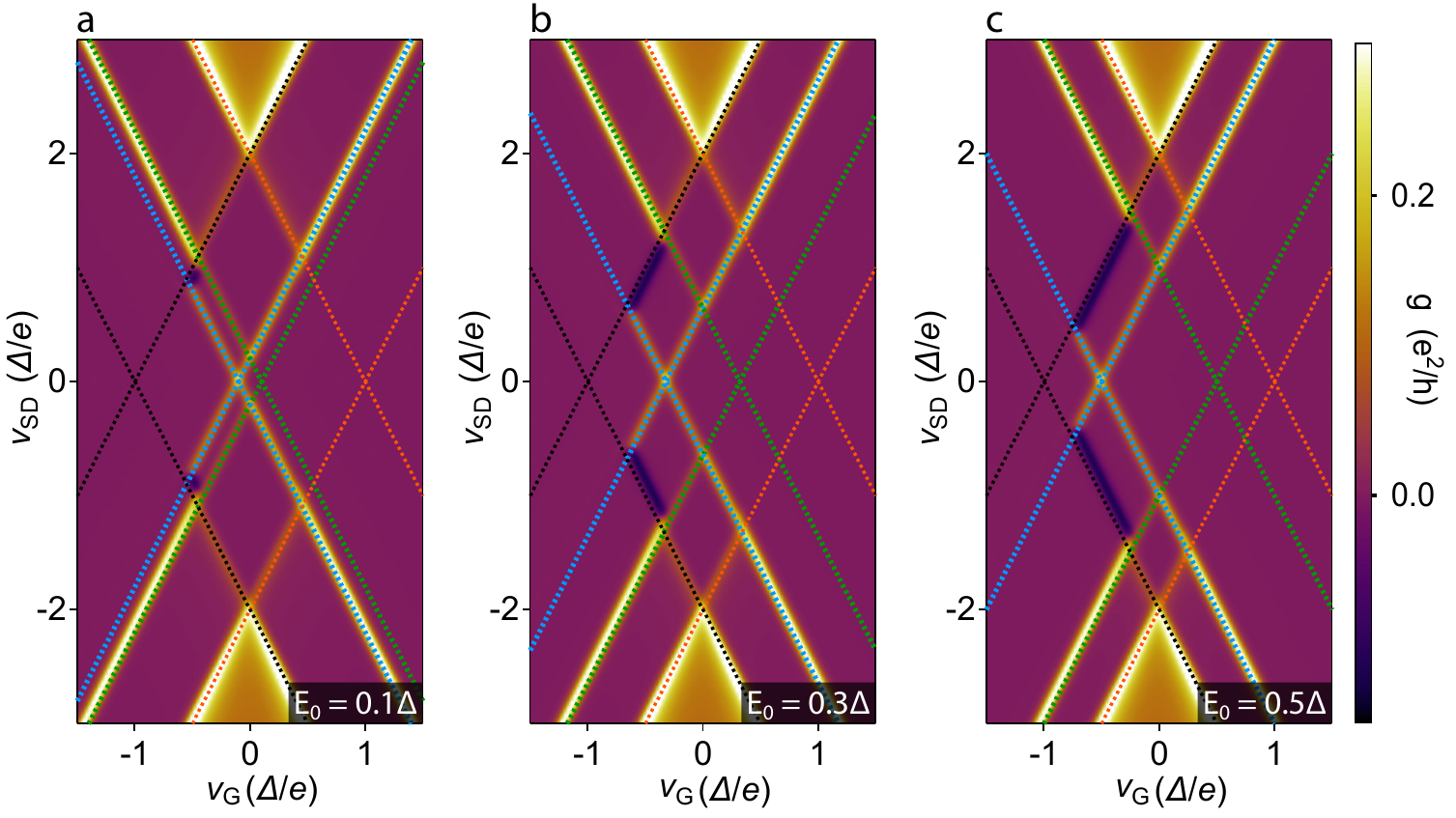}
\caption{\label{FigS3} \textbf{Interpreting conductance thresholds.}
\textbf{a},
Calculated conductance $g$ versus $v_\mathrm{SD}$ and $v_\mathrm{G}$.
$E_0=0.1\Delta$, all other model parameters same as main text.
Dotted lines are
$v_\mathrm{SD}/2 = \pm (v_\mathrm{G} + \Delta/e)$ [black],
$v_\mathrm{SD}/2 = \pm (v_\mathrm{G} + E_0/e)$ [blue],
$v_\mathrm{SD}/2 = \pm (v_\mathrm{G} - E_0/e)$ [green],
$v_\mathrm{SD}/2 = \pm (v_\mathrm{G} - \Delta/e)$ [red].
\textbf{b}, $E_0=0.32\Delta$, all model parameters same as main text.
\textbf{c}, $E_0=0.5\Delta$, all other model parameters same as main text.
Color scale shared across all plots.
}
\end{center}
\end{figure}

\subsection{ \ref{sec:model}. Derivation of transport and thermal model}
This section gives a detailed derivation of the transport and thermal model used in the main text.

To describe the electron transport through a metallic superconducting quantum dot we consider the following model:
\begin{equation}\label{ham}
H=H_{\LR}+H_{\D}+H_{\T},
\end{equation}
where the Hamiltonian
\begin{equation}
\label{hamLR}H_{\LR}=\sum_{\alpha \nu\s}\left(\ve_{\alpha\nu\s}^{{}}-\mu_{\alpha}^{{}}\right)\cd_{\alpha\nu\s}\can_{\alpha\nu\s}
\end{equation}
describes the normal metallic leads with $\cd_{\alpha\nu\s}$ being an electron creation operator in the lead $\alpha\in\{\mrL,\mrR\}$, with an orbital quantum number $\nu$ and spin $\s\in\{\up,\down\}$. The leads have chemical potentials given by $\mu_{\alpha}=\pm V/2$, where $V$ denotes symmetrically applied bias. For the semiconductor-superconductor hybrid quantum dot, we use a simplified model consisting of a Bardeen-Cooper-Schrieffer (BCS)\cite{Bardeen1957} continuum and an Andreev bound state for fixed number of particles \cite{Tinkham1972}:
\begin{equation}
\label{hamDSI}
H_{\D}=\sum_{\s,p=e,h}\Big[E_{0}\gamd_{0\s,p}\gaman_{0\s,p}
+\sum_{n}E_{n}^{{}}\gamd_{n\s,p}\gaman_{n\s,p}\Big]+E_{N}^{c},
\end{equation}
\begin{equation}
\quad E_{N}^{c}=U(N-N_g)^2, \quad N=\Big[\sum_{\s}\dd_{0\s}\dan_{0\s}+\sum_{n}\dd_{n\s}\dan_{n\s}\Big],
\end{equation}
where $\dd_{n\s}$ creates an electron in the continuum with quantum number $n$ (e.g. momentum of electrons on the dot), and $\dd_{0\s}$ denotes electron creation in a localized level, which gives rise to a subgap state in the BCS spectrum. The charging effects on the quantum dot are described by constant interaction model given by the term $E_{N}^{c}$, where the charging energy is given by $E_{C}=2U$ and the number of electrons on the dot is controlled by a gate voltage $V_{g}$, which is parameterized by dimensionless number $N_{g}=V_{g}/E_{C}$. The operator $N$ gives the total number of electrons on the dot. With superconducting pairing, the dot Hamiltonian is diagonal in the basis of the quasiparticle operators $\gamd$, which are given by \cite{Josephson1962,Bardeen1962,Tinkham2004}
\begin{subequations}
\begin{align}
&\dan_{n\s}=u_{n}\gaman_{n\s,e}+\s v_{n}\gamd_{-n\bar{\s},h},\\
&\gaman_{n\s,e}=u_{n}\dan_{n\s}-\s v_{n}\dd_{-n\bar{\s}}S,\\
&\gaman_{n\s,h}=u_{n}S^{\dag}\dan_{n\s}-\s v_{n}\dd_{-n\bar{\s}},\\
&\gamd_{n\s,e}=S^{\dag}\gamd_{n\s,h}, \quad
\gaman_{n\s,h}=S^{\dag}\gaman_{n\s,e},
\end{align}
\end{subequations}
with $S^{\dag}$ denoting the Cooper pair creation operator and $\gamd_{n\s,e/h}$ denoting the quasiparticle creation operator, which adds an electron/hole to the system. Here a state with quantum numbers $-ns$ is the time-reversed partner of a state with quantum numbers $ns$. The quasiparticle excitation energy $E_{n}$ and the BCS coherence factors $u_{n}, \ v_{n}$ are given in terms of superconducting gap $\absDelta$ and electron dispersion on the dot $\ve_{n}$ as
\begin{equation}\label{qpeEacf}
E_{n}=\sqrt{\ve_{n}^2+\absDelta^2}, \quad
u_{n}=\sqrt{\frac{1}{2}\left(1+\frac{\ve_{n}}{E_{n}}\right)}, \quad
v_{n}=\sqrt{\frac{1}{2}\left(1-\frac{\ve_{n}}{E_{n}}\right)}.
\end{equation}
Similarly, the subgap state operator is expressed as
\begin{equation}
\dan_{0\s}=u_{0}\gaman_{0\s,e}+\s v_{0}\gamd_{0\bar{\s},h},
\end{equation}
with $u_{0}$, $v_{0}$ being model dependent coherence factors, which we set to $u_{0}=v_{0}=1/\sqrt{2}$ for simplicity.
Lastly, the electrons in the leads and in the dot are coupled by the tunneling
\begin{equation}\label{hamTMI}
\begin{aligned}
H_{\T}&=
\sum_{\substack{\alpha\nu\s}}\Big[
t_{\mrS,\alpha}\cd_{\alpha\nu\s}\dan_{0\s}
+t_{\mrS,\alpha}^{*}\dd_{0\s}\can_{\alpha\nu\s}
+\sum_{n}
\left(t_{\mrC,\alpha}\cd_{\alpha\nu\s}\dan_{n\s}
+t_{\mrC,\alpha}^{*}\dd_{n\s}\can_{\alpha\nu\s}\right)\Big] \\
&\begin{aligned}=
  \sum_{\substack{\alpha\nu\s}}\Big[&t_{\mrS,\alpha}\cd_{\alpha\nu\s}(u_{0}\gaman_{0\s,e}+\s v_{0}\gamd_{0\bar{\s},h})
  +t_{\mrS,\alpha}^{*}(u_{0}^{*}\gamd_{0\s,e}+\s v_{0}^{*}\gaman_{0\bar{\s},h})\can_{\alpha\nu\s} \\
  +\sum_{n}\big\{&t_{\mrC,\alpha}\cd_{\alpha\nu\s}(u_{n}\gaman_{n\s,e}+\s v_{n}\gamd_{-n\bar{\s},h})
  +t_{\mrC,\alpha}^{*}(u_{n}^{*}\gamd_{n\s,e}+\s v_{n}^{*}\gaman_{-n\bar{\s},h})\can_{\alpha\nu\s}\big\}\Big],
 \end{aligned}
\end{aligned}
\end{equation}
where $t_{\mrC,\alpha}$ gives the tunneling amplitude to the continuum and $t_{\mrS,\alpha}$ gives the tunneling amplitude to the subgap state.

\subsubsection{Thermodynamics of the even/odd effect}

We now present the free energy difference between the superconducting metallic island having even or odd number of electrons. The parity of the number of quasiparticles has to be equal to the parity of the number of electrons $N$ on the island. The free energy difference $\delta{F}$ between the odd and even occupation is expressed as \cite{Tuominen1992,Lafarge1993}
\begin{equation}\label{dfoddeven}
\delta{F}=F_{\mrOdd}-F_{\mrEven}=-\frac{1}{\beta}\ln\left(\frac{\mcZ_{\mrOdd}}{\mcZ_{\mrEven}}\right),
\end{equation}
in terms of the partition functions for different parities
\begin{equation}\label{zevenodd}
2\mcZ_{\mrOdd/\mrEven}=\prod_{n,\s}(1+\e^{-\beta E_{n}})\mp\prod_{n,\s}(1-\e^{-\beta E_{n}}).
\end{equation}
where $\beta=1/k_\mathrm{B}T$ denotes the inverse temperature of the island. For a sufficiently large island the single particle spectrum can be described by the spectrum of a grounded superconductor where the single particle spectrum $E_{n}$ is given by Eq.~\eqref{qpeEacf}.

\textit{Without a subgap state} the free energy difference Eq.~\eqref{dfoddeven} is expressed as
\begin{equation}\label{dfbcs}
\delta{F}_{\mathrm{BCS}}
=-k_\mathrm{B}T\ln\tanh\left[\frac{1}{2}\sum_{n,\s}\ln\coth\left(\frac{\beta E_n}{2}\right)\right]
=-\frac{1}{\beta}\ln\tanh\int_{\absDelta}^{+\infty}\dif{E}\rho_{\mathrm{BCS}}(E)\ln\coth\left(\frac{\beta E}{2}\right),
\end{equation}
where $\rho_{\mathrm{BCS}}(E)$ is the BCS density of states for quasiparticles on the island given by
\begin{equation}
\rho_{\mathrm{BCS}}(E)=\frac{\rho_\mathrm{D}E}{\sqrt{E^2-\absDelta^2}},
\end{equation}
with $\rho_\mathrm{D}=\rho_{\mathrm{Al}}V$ denoting the normal state density of states, including spin, and  $\rho_{\mathrm{Al}}$ is aluminum density of states per volume, and $V$ is the volume of the island. For small temperatures $\beta\Delta\gg1$, the free energy difference \eqref{dfbcs} can be approximated as
\begin{equation}
\delta{F}_{\mathrm{BCS}}
\approx-k_\mathrm{B}T\ln\tanh\left[2\int_{\absDelta}^{+\infty}\dif{E}\rho_{\mathrm{BCS}}(E)\e^{-\beta E}\right]
=-k_\mathrm{B}T\ln\tanh\left[N_\mathrm{eff} e^{-\beta \Delta}\right]\approx \Delta - k_\mathrm{B}T \ln( N_\mathrm{eff} ),
\end{equation}
where an effective number of quasiparticle states $N_\mathrm{eff}$ is given by
\begin{equation}\label{eq:neff1}
N_\mathrm{eff}=2 \int_{\absDelta}^{+\infty}\dif{E} \rho_{\mathrm{BCS}}(E) e^{-\beta( E - \Delta )} = 2 \rho_\mathrm{D} \Delta e^{ \beta \Delta } K_{1}( \beta \Delta )
\approx \rho_\mathrm{D} \sqrt{ 2 \pi k_\mathrm{B} T \Delta }
\end{equation}
and $K_{\nu}(x)$ denotes the modified Bessel function of the second kind.

\textit{With a subgap state} the free energy difference Eq.~\eqref{dfoddeven} acquires an additional term and one gets
\begin{equation}
\delta{F}_{\mathrm{ABS}}
=-k_\mathrm{B}T\ln\tanh\left[\int_{\absDelta}^{+\infty}\dif{E}\rho_{\mathrm{BCS}}(E)\ln\coth\left(\frac{\beta E}{2}\right)+\ln\coth (\beta E_0/2)\right].\label{eq:dFABS}
\end{equation}
See also Eq.~\eqref{eq:dfbs} where the approximate expression for the first term in used.

In the main text we discuss how the low-temperature data deviates from the above Andreev-bound-state model in terms of a life-time broadening of the subgap state. This is done by including a phenomenological broadening with width $\gamma$ into the subgap density of states, which then gives the free energy difference
\begin{equation}
\label{eq:dfint}
\delta{F}_{\mathrm{ABS-LB}}=-k_\mathrm{B}T\ln\tanh\Bigg[\int_{\absDelta}^{+\infty}\dif{E}\rho_{\mathrm{BCS}}(E)\ln\coth\left(\frac{\beta E}{2}\right)
+\frac{1}{2}\sum_{\substack{\tau=\pm1 \\ s=\up,\down}}\int_{0}^{+\infty}\frac{\dif{\omega}}{2\pi}\frac{\gamma \ln\coth\left(\frac{\beta\omega}{2}\right)}{(\omega-\tau E_0)^2+(\gamma/2)^2}\Bigg].
\end{equation}
In the kinetic equation calculation presented below, the equilibrium distributions of quasiparticle in the continuum with an even or odd number of quasiparticles are needed. Since we will assume that the particles occupying the continuum are effectively equilibrated, we find the distribution functions by modifying the usual Fermi-Dirac distribution function as
\begin{equation}\label{qpdist}
f_{P}(E)=\frac{1}{\e^{\beta E}(\mcZ_{P}/\mcZ_{\bar{P}})+1}\quad\rightarrow\quad
\left\{\begin{aligned}
&f_{\mrEven}(E)=\frac{1}{\e^{\beta(E+\delta{F}_\mathrm{BCS})}+1},\\
&f_{\mrOdd}(E)=\frac{1}{\e^{\beta(E-\delta{F}_\mathrm{BCS})}+1},
\end{aligned}\right.
\end{equation}
where $P\in\{\mrEven,\mrOdd\}$, and $\bar{P}$ represents the opposite of $P$.

\subsubsection{Number of quasiparticles}
Using the above results, we derive a simple expression for the number of quasiparticles in the absence of a bound state.
At low temperature, when $\delta{F}_{\mathrm{BCS}} = \Delta - k_\mathrm{B} T \ln( N_\mathrm{eff} )$, the distribution functions take the form
\begin{eqnarray}
f_{\mrEven} &=& N_\mathrm{eff} e^{-\beta( E + \Delta ) }, \\
f_{\mrOdd} &=& \frac{1}{N_\mathrm{eff}} e^{-\beta( E - \Delta ) },
\end{eqnarray}
where $N_\mathrm{eff}$ is given by Eq.~(\ref{eq:neff1}).

The number of quasiparticles in each parity state can then be calculated using
\begin{equation}
N_{P} = 2 \int_{\absDelta}^{+\infty}\dif{E} \rho_\mathrm{BCS}(E)  f_{P}(E).
\end{equation}
Substituting the above expression for $f_{\mrOdd}$ gives $N_{\mrOdd} = 1$, as expected.
Substituting $f_{\mrEven}$ gives the quasiparticle number
\begin{eqnarray}
N_{\mrEven} &=& 2 \int_{\absDelta}^{+\infty}\dif{E} \rho_\mathrm{BCS}(E) N_\mathrm{eff} e^{-\beta( E + \Delta ) } \notag\\
&=& N_\mathrm{eff} e^{-2 \beta \Delta } \int_{\absDelta}^{+\infty}\dif{E}~2\rho_\mathrm{BCS}(E)  e^{-\beta( E - \Delta ) } \notag\\
&=& \left( N_\mathrm{eff} e^{ - \beta \Delta} \right)^2.
\end{eqnarray}
Because of the large charging energy $N_{\mrEven}$ is the square of the bulk value $N_\mathrm{eff} e^{ - \beta \Delta}$, indicating that quasiparticles must be created in pairs.

\subsubsection{Incoherent transport}

We now calculate the current through the quantum dot by a set of master equations, with transition rates calculated by Fermi's golden rule, which is valid in the weak tunneling regime or the so-called sequential tunneling regime.

According to the Fermi's golden rule the transition rate from the initial state $i$ to the final state $j$ caused by a perturbation $V$ is given by \cite{Bruus2004}:
\begin{equation}\label{fgr}
\Gamma_{fi}=2\pi\abs{\bra{f}V\ket{i}}^2\delta(E_f-E_i).
\end{equation}
In our case we are interested in the transitions caused by the tunneling Hamiltonian \eqref{hamTMI} between the states $\ket{\m{D}}\ket{\LR}$ and $\ket{\m{D'}}\ket{\LR'}$:
\begin{equation}\label{gma1}
\Gamma_{\m{D}'\m{D},\LR'\LR}=2\pi\abs{\bra{\LR'}\bra{\m{D}'}H_\T\ket{\m{D}}\ket{\LR}}^2\delta(E_{\m{D}'}-E_{\m{D}}+E_{\LR'}-E_{\LR}),
\end{equation}
where the many-body eigenstates of the lead Hamiltonian $H_{\LR}$ are denoted as $\ket{\LR}$ and of the dot Hamiltonian $H_{\D}$ are denoted as $\ket{\m{D}}$. Also $E_{\m{D}}$ gives the energy of the state $\ket{\m{D}}$ and $E_{\LR}$ gives the energy of the state $\ket{\LR}$.
So we see that we need the following matrix element
\begin{equation}\label{mesqSI}
\begin{aligned}
\abs{\bra{\LR'}\bra{\m{D}'}H_\T\ket{\m{D}}\ket{\LR}}^2=\sum_{\substack{\alpha\nu\s n}}\abs{t_{\alpha}}^2\Big\{
&\abs{u_{n}}^2\abs{\bra{\LR'}\cd_{\alpha\nu\s}\ket{\LR}}^2
 \abs{\bra{\m{D}'}\gaman_{n\s,e}\ket{\m{D}}}^2\\
+&\abs{v_{n}}^2\abs{\bra{\LR'}\cd_{\alpha\nu\s}\ket{\LR}}^2
  \abs{\bra{\m{D}'}\gamd_{-n\bar{\s},h}\ket{\m{D}}}^2\\
+&\abs{u_{n}}^2\abs{\bra{\LR'}\can_{\alpha\nu\s}\ket{\LR}}^2
  \abs{\bra{\m{D}'}\gamd_{n\s,e}\ket{\m{D}}}^2\\
+&\abs{v_{n}}^2\abs{\bra{\LR'}\can_{\alpha\nu\s}\ket{\LR}}^2
  \abs{\bra{\m{D}'}\gaman_{-n\bar{\s},h}\ket{\m{D}}}^2\Big\}.
\end{aligned}
\end{equation}
Note that we have not written out the terms with the subgap state, but they have an analogous structure.

In order to form the tunneling rates we need to specify what kind of states $\ket{\m{D}}$ on the metallic superconducting quantum dot we consider and what kind of (thermal) averaging procedure for these states we employ. We classify the states $\ket{\m{D}}$ according to the number of electrons $N$, the number of excitations $N_{S}$ in the continuum, quantum number $l$ labeling the configuration of the excitations in the state, and the occupancy of the subgap state $x\in\{0,\up,\down,2\}$, i.e.,
\begin{equation}
\ket{\m{D}}\equiv\ket{N,N_{S},l,x}.
\end{equation}
The above states have the energy
\begin{equation}
E_{N,N_{S},l,x}=E_{N}^{c}+E_{x}+\sum_{n\in l}^{N_{S}}E_{n}.
\end{equation}
If the number of electrons $N$ is even/odd then the total number of excitations $N_{\mathrm{tot},S}$ also has to be even/ood. This means that the number of excitations in the continuum $N_{S}$ depends on the subgap state occupancy $x$ and the charge on the dot $N$, i.e.,
\begin{equation}\label{presNS}
N_{S}\rightarrow
\left\{\begin{aligned}
&\text{even for $N$ even and $x\in\{0,2\}$},\\
&\text{odd for $N$ even and $x\in\{\up,\down\}$},\\
&\text{odd for $N$ odd and $x\in\{0,2\}$},\\
&\text{even for $N$ odd and $x\in\{\up,\down\}$}.
\end{aligned}
\right.
\end{equation}

Now we want to thermally average over all quasiparticle states $N_{S}$  of the continuum and their configurations $l$. The thermal averaging assumption is valid if the relaxation rate of quasiparticles is much faster than the rate of the tunneling events between the leads and the island. From the expression (\ref{mesqSI}) we see that we need to consider the thermal averages of the following expressions for the dot
\begin{subequations}\label{qpediNS}
\begin{equation}
f_{\mrS}(E_{n},N,x)=\sum_{l,N_{S}}W_{N,N_{S},l,x}\abs{\bra{N,N_{S},l,x}\gamd_{n\s,p}\gaman_{n\s,p}\ket{N,N_{S},l,x}}^2,
\end{equation}
\begin{equation}
\bar{f}_{\mrS}(E_{n},N,x)=\sum_{l,N_{S}}W_{N,N_{S},l,x}\abs{\bra{N,N_{S},l,x}\gaman_{n\s,p}\gamd_{n\s,p}\ket{N,N_{S},l,x}}^2
=1-f_{\mrS}(E_{n},N,x),
\end{equation}
\end{subequations}
where $p=e,h$. Here $W_{N,N_{S},l}$ denotes a thermal distribution for which we have
\begin{equation}\label{probSMI}
W_{N,N_{S},l,x}=\frac{1}{\mcZ_{x}}\e^{-\beta E_{N,N_{S},l,x}}, \quad \mcZ_{x}=\sum_{N_{S},l}W_{N,N_{S},l,x}.
\end{equation}
By using Eq. \eqref{qpdist} and following the prescription \eqref{presNS}, we get the distributions
\begin{equation}
f_{\mrS}(E,N,x)=\left\{\begin{aligned}
&f_{\mrEven}(E), \quad \text{for $N$ even and $x=0,2$},\\
&f_{\mrOdd}(E), \quad \text{for $N$ even and $x=\up,\down$},\\
&f_{\mrOdd}(E), \quad \text{for $N$ odd and $x=0,2$},\\
&f_{\mrEven}(E), \quad \text{for $N$ odd and $x=\up,\down$}.
\end{aligned}\right.
\end{equation}

After thermally averaging over the source-drain lead states $\ket{\LR}$, using grand-canonical ensemble, and the continuum states of the dot, we obtain the following tunneling rates from and to the continuum of the dot
\begin{equation}\label{gamSI}
\begin{aligned}
\Gamma_{\substack{N+\chi\leftarrow N \\ N_S+\chi'\leftarrow N_S}}^{\alpha}&\approx \gamma_{\alpha}
\int_{\abs{\Delta}}^{+\infty}\frac{E\dif E}{\sqrt{E^2-\abs{\Delta}^2}}
f_{\mrN}(E^{c}_{N+\chi}-E^{c}_{N}+\chi'E-\chi\mu_{\alpha})f_{\mrS,-\chi'}(E,N,x),\\
&\text{with} \quad f_{\mrS,+}(E,N,x)=f_{\mrS}(E,N,x), \quad f_{\mrS,-}(E,N,x)=\bar{f}_{\mrS}(E,N,x).
\end{aligned}
\end{equation}
Here $\gamma_{\alpha}=2\times 2\pi\times \rho_\mathrm{D}\rho_{\alpha}\abs{t_{\mrC,\alpha}}^2$ with $\rho_{\alpha}$ denoting density of states of the normal leads.
The continuum coupling $\gamma_{\alpha}=2\times 2\pi\times \rho_\mathrm{D}\rho_{\alpha}\abs{t_{\mrC,\alpha}}^2$ is related to the normal-state conductance by $g_\mathrm{Al}=(\pi/2) (e^2/h) \gamma_{\alpha}$.
Note that we have set the chemical potential of the metallic superconducting dot at zero $\mu_{D}=0$ in order not to complicate the calculations, and also used that $\ve_{n}=\ve_{-n}$.

Additionally, there are tunneling rates from and to the subgap state. When the starting state has no quasiparticles in the subgap state, i.e., $\ket{0}$, we get the following rates
\begin{subequations}\label{gamSGS0}
\begin{align}
&\begin{aligned}
\Gamma_{\substack{N-1\leftarrow N \\ \s\leftarrow 0}}^{\alpha}
&\approx \Gamma_{\alpha}v_{0}^2 [1-f_{\mrN}(E^{c}_{N}-E_{\s}-E^{c}_{N-1}-\mu_{\alpha})],
\end{aligned}\\
&\begin{aligned}
\Gamma_{\substack{N+1\leftarrow N \\ \s\leftarrow 0}}^{\alpha}
&\approx \Gamma_{\alpha} u_{0}^2 f_{\mrN}(E^{c}_{N+1}+E_{\s}-E^{c}_{N}-\mu_{\alpha}).
\end{aligned}
\end{align}
\end{subequations}
For the state with single quasiparticle $\ket{s}$ we get
\begin{subequations}\label{gamSGSs}
\begin{align}
&\begin{aligned}
\Gamma_{\substack{N-1\leftarrow N \\ 0\leftarrow \s}}^{\alpha}
&\approx \Gamma_{\alpha}u_{0}^2[1-f_{\mrN}(E^{c}_{N}+E_{\s}-E^{c}_{N-1}-\mu_{\alpha})],
\end{aligned}\\
&\begin{aligned}
\Gamma_{\substack{N-1\leftarrow N \\ 2\leftarrow \s}}^{\alpha}
&\approx \Gamma_{\alpha}v_{0}^2 [1-f_{\mrN}(E^{c}_{N}-E_{\bar{\s}}-E^{c}_{N-1}-\mu_{\alpha})],
\end{aligned}\\
&\begin{aligned}
\Gamma_{\substack{N+1\leftarrow N \\ 2\leftarrow \s}}^{\alpha}
&\approx \Gamma_{\alpha} u_{0}^2 f_{\mrN}(E^{c}_{N+1}+E_{\bar{\s}}-E^{c}_{N}-\mu_{\alpha}),
\end{aligned}\\
&\begin{aligned}
\Gamma_{\substack{N+1\leftarrow N \\ 0\leftarrow \s}}^{\alpha}
&\approx \Gamma_{\alpha}v_{0}^2 f_{\mrN}(E^{c}_{N+1}-E_{\s}-E^{c}_{N}-\mu_{\alpha}),
\end{aligned}
\end{align}
\end{subequations}
and for the state with two quasiparticles $\ket{2}$ we get
\begin{subequations}\label{gamSGS2}
\begin{align}
&\begin{aligned}
\Gamma_{\substack{N-1\leftarrow N \\ \s\leftarrow 2}}^{\alpha}
&\approx \Gamma_{\alpha}u_{0}^2[1-f_{\mrN}(E^{c}_{N}+E_{\bar{s}}-E^{c}_{N-1}-\mu_{\alpha})],
\end{aligned}\\
&\begin{aligned}
\Gamma_{\substack{N+1\leftarrow N \\ \s\leftarrow 2}}^{\alpha}
&\approx \Gamma_{\alpha}v_{0}^2 f_{\mrN}(E^{c}_{N+1}-E_{\bar{s}}-E^{c}_{N}-\mu_{\alpha}).
\end{aligned}
\end{align}
\end{subequations}
Here $\Gamma_{\alpha}=2\pi\rho_{\alpha}\abs{t_{\mrS,\alpha}}^2$, and $s\in\{\up,\down\}$ with $\bar{s}$ denoting the opposite of $s$. We include the relaxation from the continuum to the subgap state by introducing the following rates within the same charge state
\begin{equation}
\Gamma_{\substack{N_{\mrOdd}\leftarrow N_{\mrOdd} \\ 0\leftarrow \s}}=
\Gamma_{\substack{N_{\mrEven}\leftarrow N_{\mrEven} \\ 2\leftarrow \s}}=\Gamma_{\mathrm{relax}}.
\end{equation}

Now we want to find the current through the superconducting metallic quantum dot. To do this we need to obtain the occupation probabilities $P_{N,x}$ of the states described by a number of electrons $N$ on the dot and the occupancy of the subgap state $x$. We write the following steady state Pauli master equation for the probabilities $P_{N, x}$
\begin{equation}\label{ameqSI}
\frac{\dif}{\dif t} P_{N,x}
=-\sum_{N'x'}\Gamma_{\substack{N'\leftarrow N \\ x'\leftarrow x}}P_{N,x}
 +\sum_{N'x'}\Gamma_{\substack{N\leftarrow N' \\ x\leftarrow x'}}P_{N',x'}=0,
\end{equation}
with the condition
\begin{equation}
\sum_{N,x} P_{N,x}=1.
\end{equation}
The rates entering in (\ref{ameqSI}) are given by
\begin{equation}
\Gamma_{\substack{N'\leftarrow N \\ x'\leftarrow x}}=\Gamma_{\substack{N'\leftarrow N \\ x'\leftarrow x}}^{L}+\Gamma_{\substack{N'\leftarrow N \\
x'\leftarrow x}}^{R},
\end{equation}
and for $x=x'$ we have
\begin{equation}
\Gamma_{\substack{N'\leftarrow N \\ x\leftarrow x}}^{\alpha}=
\Gamma_{\substack{N'\leftarrow N \\ N_S-1\leftarrow N_S}}^{\alpha,x}+\Gamma_{\substack{N'\leftarrow N \\ N_S+1\leftarrow N_S}}^{\alpha,x}.
\end{equation}
When the occupation probabilities $P_{N,x}$ are obtained, the current through the quantum dot can be written as
\begin{equation}\label{crteq}
I_{\mathrm{seq}}=(-e)\sum_{N,xx'}\left(
\Gamma^{L}_{\substack{N+1\leftarrow N \\ x'\leftarrow x}}
-\Gamma^{L}_{\substack{N-1\leftarrow N \\ x'\leftarrow x}}
\right)P_{N,x}.
\end{equation}

These formulae are then solved numerically to produce the plots in Fig. S1 and Fig. 2b in the main text.

\subsection{ \ref{sec:dFapprox}. Comparison of free energy approximations}
This section gives examples of the free energy difference, $F_\mathrm{o}-F_\mathrm{e}$, calculated under different approximations, considering the case without broadening $\gamma=0$ and without an applied field $B=0$.
Under these conditions the free energy difference is given by Eq.~\eqref{eq:dFABS}. When $\beta \Delta > 1$ the approximation $\ln \coth ( \beta E / 2 ) \approx 2 e^{-\beta E}$ can be used for the first term.
Applying the identity $\int_{\absDelta}^{+\infty}\dif{E}\rho_{\mathrm{BCS}}(E) e^{- \beta E } = \rho_\mathrm{Al} V \Delta K_1( \beta \Delta )$ then gives
\begin{equation}
\label{eq:dfbs}
F_\mathrm{o}-F_\mathrm{e} \approx -k_\mathrm{B}T\ln\tanh\Bigg[ 2 \rho_\mathrm{Al} V \Delta K_{1}( \beta \Delta ) + \ln\coth\left(\frac{\beta E_0}{2}\right)\Bigg],
\end{equation}
where $K_1( x )$ is a Bessel function of the second kind.
In the very low temperature limit $\beta \Delta \gg 1$, $\beta E_0 > 1$ the approximations $K_1( \beta \Delta ) \approx \sqrt{ \pi / (2 \beta \Delta) } e^{-\beta \Delta}$, $\ln \coth ( \beta E_0 / 2 ) \approx 2 e^{-\beta E_0}$, and $\tanh( x ) \approx x$ can be used, giving
\begin{equation}
\label{eq:dfeasy}
F_\mathrm{o}-F_\mathrm{e} \approx k_\mathrm{B}T\ln \Bigg[ N_\mathrm{eff} e^{-\beta \Delta} + 2 e^{-\beta E_0} \Bigg],
\end{equation}
where $N_\mathrm{eff}=\rho_\mathrm{Al} V \sqrt{ 2 \pi k_\mathrm{B} T \Delta} $.

\begin{figure}[h]
	\includegraphics{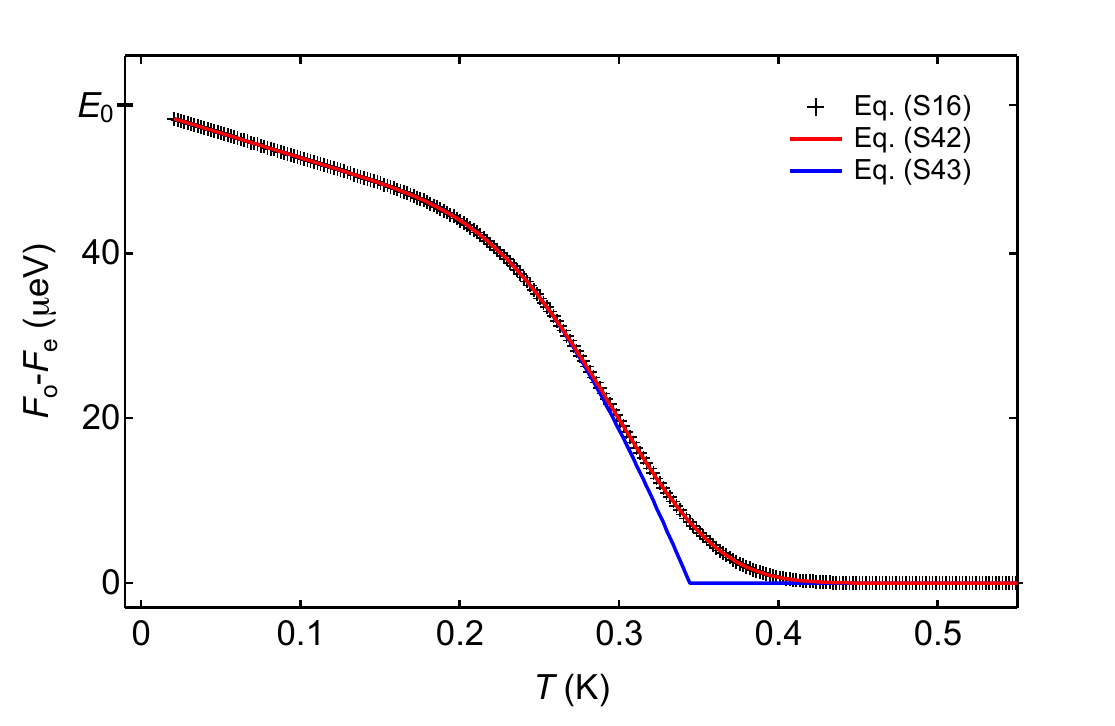}
	\caption{\textbf{Comparison of free energy approximations.} Free energy difference $F_\mathrm{o}-F_\mathrm{e}$ versus temperature $T$ for three different expressions for the free energy.
	All parameters same as main text ($\Delta=180~\mathrm{\mu e V}$, $E_0=58~\mathrm{\mu e V}$, $\gamma=0$, $B=0$).
	Black crosses are numerically exact values from Eq.~(\ref{eq:dFABS}), red line is Eq.~(\ref{eq:dfbs}), blue line is Eq.~(\ref{eq:dfeasy}).
	}
\end{figure}

Equations (\ref{eq:dfbs}) and (\ref{eq:dfeasy}) constitute two levels of accuracy at which Eq.~(\ref{eq:dFABS}) can be evaluated.
Figure~S4 compares the methods.
Equation~(\ref{eq:dfbs}) is an excellent approximation to Eq.~(\ref{eq:dFABS}) over the experimentally relevant temperature range.
Equation~(\ref{eq:dfeasy}) is poor approximation at intermediate temperatures.

\subsection{ \ref{sec:dFBS}. Effect of the bound state on the free energy }
Figure~S5 shows a comparison of the free energy difference, $F_\mathrm{o}-F_\mathrm{}$, with and without the subgap bound state. As the lowest energy unoccupied state, the bound state causes the free energy to saturate at $F_\mathrm{o}-F_\mathrm{e}=E_0$ at low temperature. It should be noted that the free energy difference with a subgap bound state was also shown in Fig. 5 of Lafarge et al. \cite{Lafarge1993}.

\begin{figure}[h]
	\includegraphics{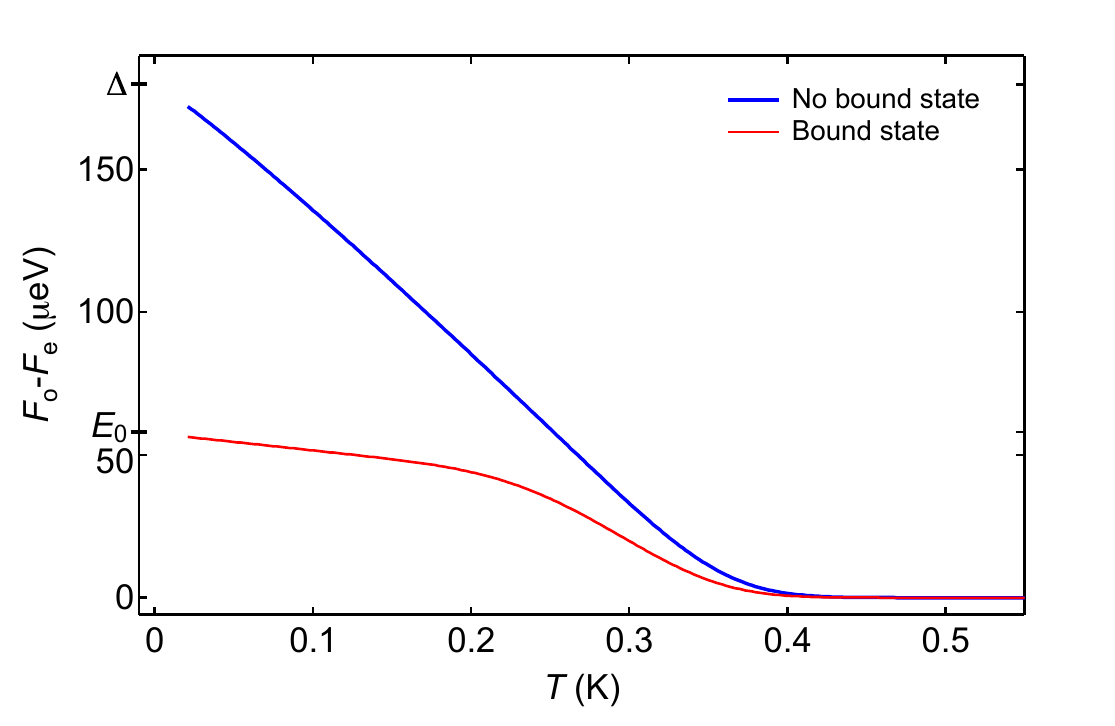}
	\label{fig:bscompare}
	\caption{\textbf{Effect of bound state on free energy.} Free energy difference $F_\mathrm{o}-F_\mathrm{e}$ versus temperature $T$ with and without the semiconductor bound state.
	All parameters the same as main text ($\Delta=180~\mathrm{\mu e V}$, $E_0=58~\mathrm{\mu e V}$, $\gamma=0$, $B=0$).
	Blue trace includes the BCS density of states only, red trace includes the BCS density of states and the discrete state.
	}
\end{figure}

\end{document}